\documentclass[12pt]{article}
\usepackage{graphicx}
\usepackage{xspace}

\usepackage[colorlinks=true, urlcolor=blue, citecolor=black]{hyperref}  

\newcommand\pubnumber{DPF2015-59}
\newcommand\pubdate{\today}

\newcommand{\MET}{\ensuremath{E_{\mathrm{T}}^{\rm{miss}}}\xspace}
\newcommand{\bbbar}{\ensuremath{b\overline{b}}\xspace}

\def\osu{Department of Physics \\
The Ohio State University, Columbus, Ohio 43210}
\def\support{\footnote{On behalf of the CMS Collaboration}}

\def\Title#1{\begin{center} {\Large #1 } \end{center}}
\def\Author#1{\begin{center}{ \sc #1} \end{center}}
\def\Address#1{\begin{center}{ \it #1} \end{center}}

\newcommand\pubblock{\rightline{\begin{tabular}{l} \pubnumber\\
         \pubdate  \end{tabular}}}
\newenvironment{Abstract}{\begin{quotation}  }{\end{quotation}}
\newenvironment{Presented}{\begin{quotation} \begin{center} 
             PRESENTED AT\end{center}\bigskip 
      \begin{center}\begin{large}}{\end{large}\end{center} \end{quotation}}
\def\Acknowledgments{\bigskip  \bigskip \begin{center} \begin{large}
             \bf ACKNOWLEDGMENTS \end{large}\end{center}}

\def\beq{\begin{equation}}
\def\eeq#1{\label{#1}\end{equation}}
\def\eeqn{\end{equation}}

\def\beqa{\begin{eqnarray}}
\def\eeqa#1{\label{#1}\end{eqnarray}}
\def\eeqan{\end{eqnarray}}

\let\bar=\overbar

\def\Dslash{\not{\hbox{\kern-4pt $D$}}}
\def\dslash{\not{\hbox{\kern-2pt $\del$}}}

\def\msb{{\bar{\ssstyle M \kern -1pt S}}}

\begin{document}
\begin{titlepage}
\pubblock

\vfill
\Title{Recent CMS searches for exotic phenomena beyond the Standard Model} 
\vfill
\Author{ H. Wells Wulsin \support}
\Address{\osu}
\vfill
\begin{Abstract}
The results of several recent CMS searches for exotic phenomena beyond the Standard Model are presented in this talk.  
Two searches look for new physics in a final state with a vector boson and missing transverse energy.  Three searches target massive resonances decaying to a Higgs boson and a vector boson.  Finally, preliminary results are presented for the first CMS search for exotic phenomena using $\sqrt{s} = 13$ TeV data, the search for dijet resonances.  
\end{Abstract}
\vfill
\begin{Presented}
DPF 2015\\
The Meeting of the American Physical Society\\
Division of Particles and Fields\\
Ann Arbor, Michigan, August 4--8, 2015\\
\end{Presented}
\vfill
\end{titlepage}
\def\thefootnote{\fnsymbol{footnote}}
\setcounter{footnote}{0}

\section{Introduction}

There are many important questions that are not answered by the Standard Model, including the nature of dark matter and an explanation for what stabilizes the Higgs mass against large quantum corrections (the hierarchy problem).  
The CMS experiment~\cite{CMSDet} at the LHC conducts a broad search program for exotic phenomena beyond the Standard Model that could resolve some of these unanswered questions~\cite{EXOPublic}.  


This talk presents the results of several recent searches for exotica phenomena, based on three signatures:   a vector boson plus missing transverse energy, a massive resonance decaying to a Higgs boson and a vector boson, and a massive resonance decaying to a pair of jets.  The searches were conducted using the $\sqrt{s} = 8$ TeV data collected in 2012, with the exception of the dijet resonance search, which has been performed with the $\sqrt{s} = 13$ TeV data recorded in 2015.  

\section{V + \MET}
If dark matter were to be produced at the LHC, it may be identified by events with a large imbalance in momentum in the transverse plane, produced by the recoil of Standard Model particles against undetected dark matter particles.  The missing transverse energy \MET is defined as the negative vector sum of the transverse momenta of all identified physics objects in an event.  Dark matter searches may look for large \MET in conjunction with a high-momentum jet, photon, $t$-quark, or other Standard Model objects.  
Two recent CMS searches pursue new physics that produces events with large \MET and a W or Z vector boson.

\subsection{W/Z(jets) + \MET}  
A search for events with large \MET and a W or Z vector boson that decays to jets~\cite{VMETHad} divides events among three categories based on the topology of jets.  In the \textit{boosted V-tagged} category, the vector boson V = W,Z produces a single fat jet, which is identified based on the jet mass and N-subjettiness, a measure of the number of subjets in a given jet.  The \textit{resolved V-tagged} category is defined as a boson that decays to two jets, which is identified as a W or Z based on a multivariate analyzer.  Finally, events that are not included in either of the V-tagged categories are placed into the \textit{monojet} category.  

\begin{figure}[!h]
\centering
\begin{minipage}{0.45\linewidth}
\centerline{\includegraphics[width=\linewidth]{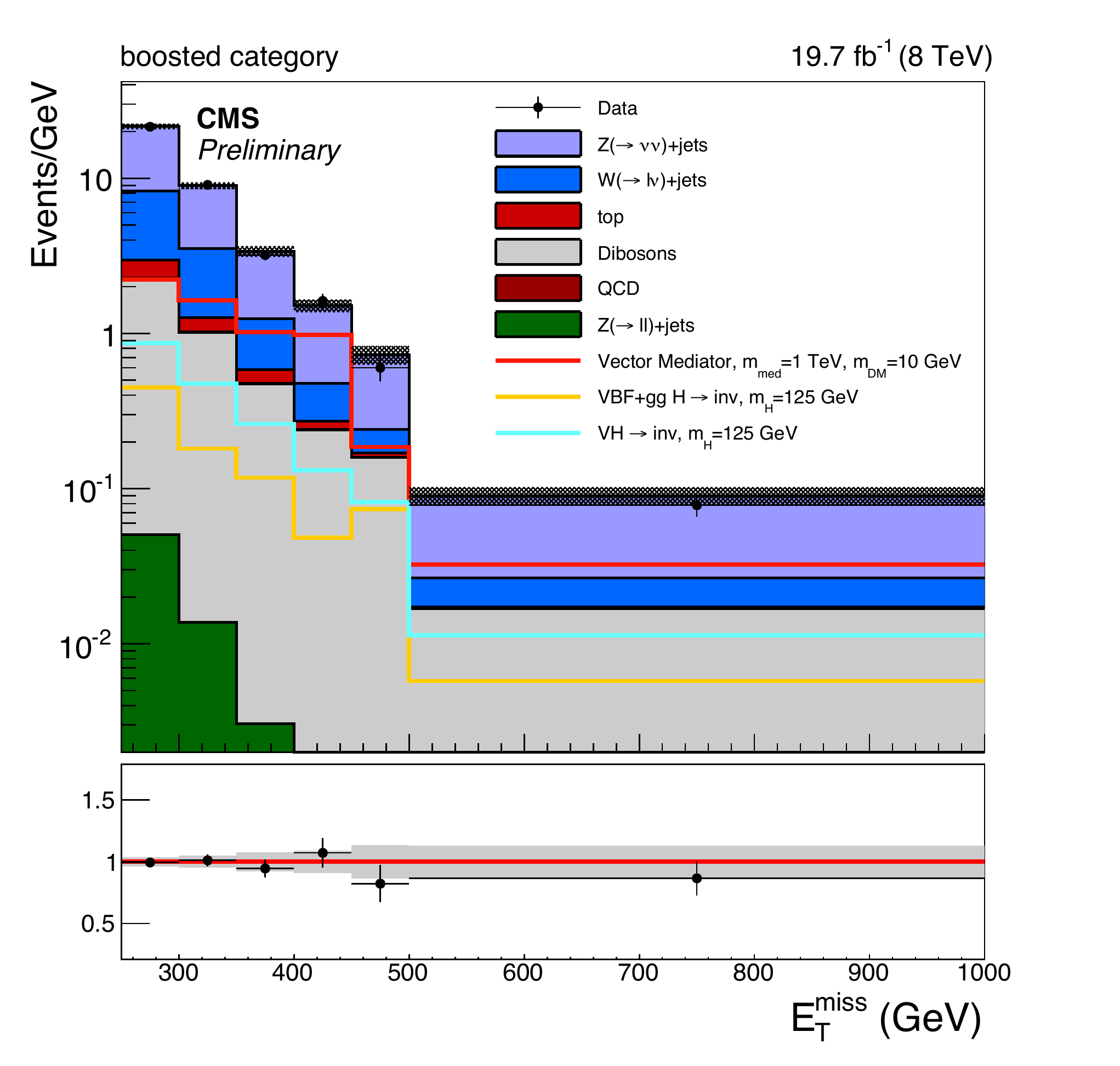}}
\end{minipage}
\hfill
\begin{minipage}{0.53\linewidth}
\centerline{\includegraphics[width=\linewidth]{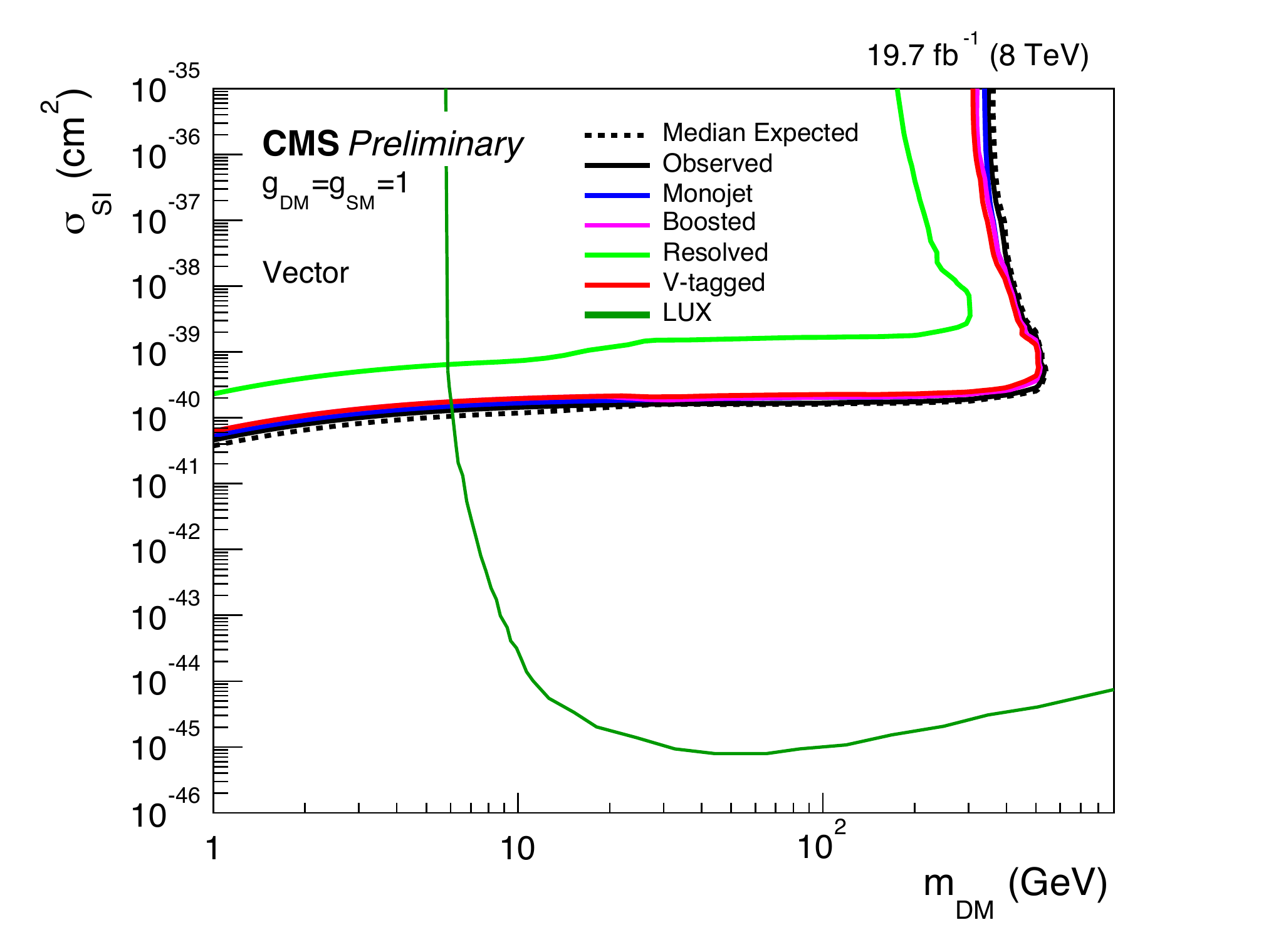}}
\end{minipage}
\caption{Left:  the post-fit distribution of \MET expected from Standard Model backgrounds and observed in data
in the signal region, for the boosted V-tagged category.  
Right:  exclusion contours in the plane of the cross section versus mass of the dark matter candidate, assuming a vector mediator. }
\label{fig:VMETHad}
\end{figure}

The background shapes and normalization for this search are taken from data control regions.  For each of the three categories, a fit is performed to the \MET distribution.  The data are found to be consistent with the predicted Standard Model background, as shown for the boosted V-tagged category in Figure~\ref{fig:VMETHad}.   

An upper limit of 53\% is placed on the branching fraction of Higgs boson decays to invisible particles.  
Limits are calculated for spin-dependent and spin-independent dark matter-nucleon scattering cross sections, for low- and high-mass mediators and compared with direct detection bounds.  
Exclusion limits in terms of the mediator mass and dark matter mass are placed on simplified models with vector, axial-vector, scalar, and pseudo scalar mediators.  The exclusion for a vector mediator is shown in Figure~\ref{fig:VMETHad}.

\subsection{$Z\rightarrow \ell\ell$ + \MET}  
\begin{figure}[!h]
\centering
\begin{minipage}{0.45\linewidth}
\centerline{\includegraphics[width=\linewidth]{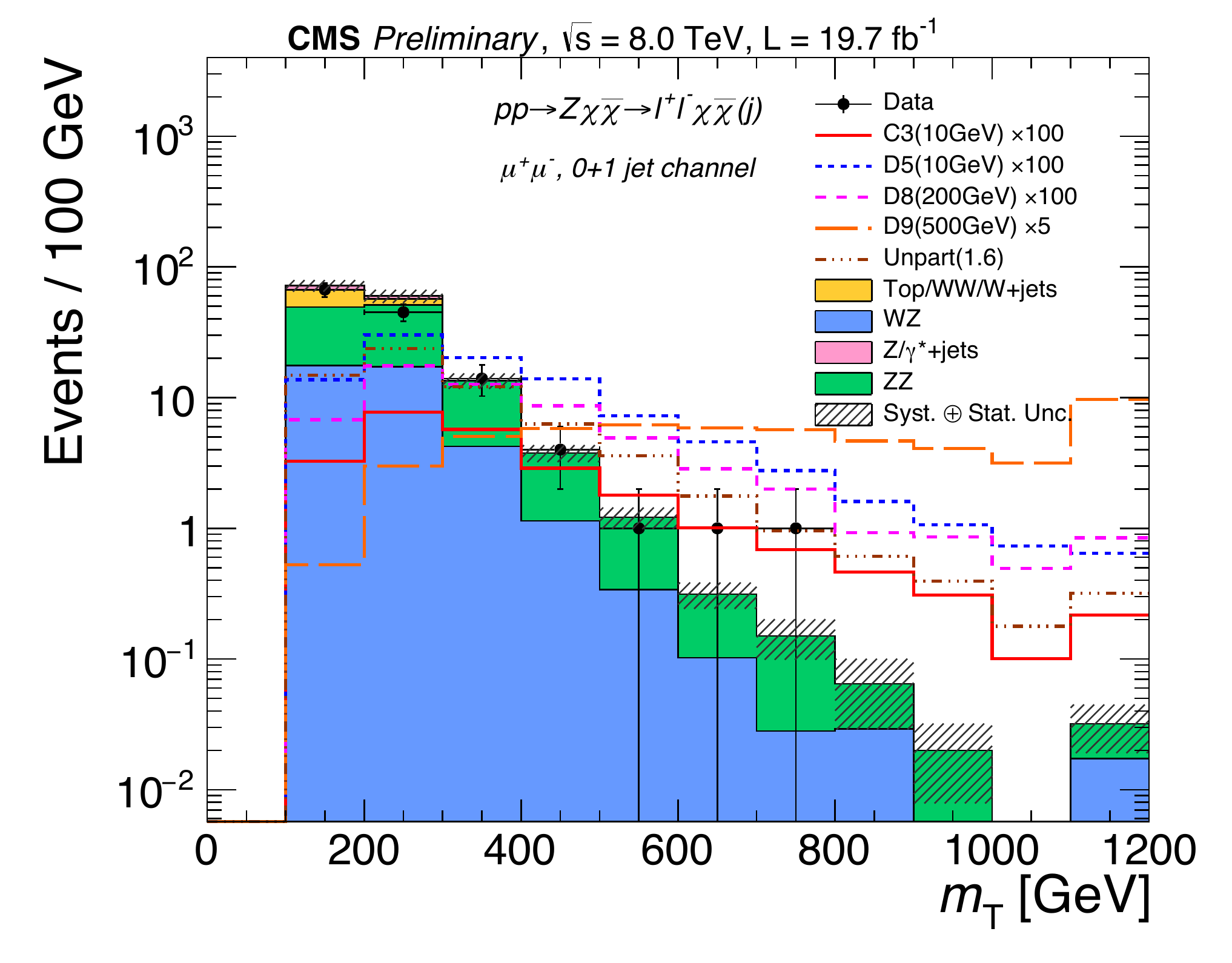}}
\end{minipage}
\hfill
\begin{minipage}{0.45\linewidth}
\centerline{\includegraphics[width=\linewidth]{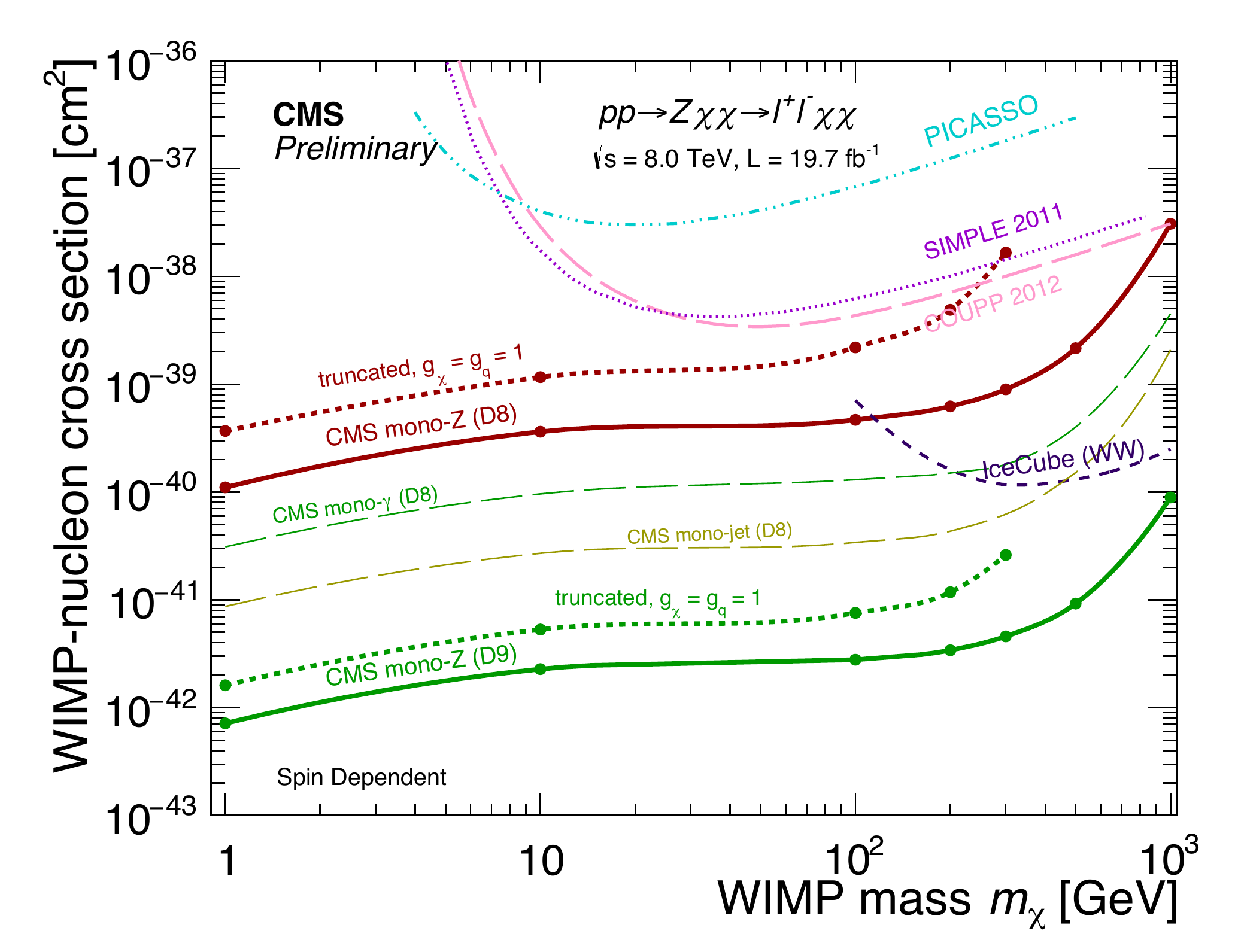}}  
\end{minipage}
\caption{ 
Left:  the distribution of the transverse mass for the final selection in the Z$\rightarrow\mu\mu$ channel.  Right:  the 90\% CL upper limits on the spin-dependent dark matter-nucleon cross section for axial-vector (D8) and tensor (D9) coupling of Dirac fermion dark matter candidates as a function of the dark matter particle mass.  Direct search results from other experiments are also shown.  }
\label{fig:VMETLep}
\end{figure}

A related search selects events with large \MET and a Z boson decaying to a pair of electrons or muons~\cite{VMETLep}.  
By triggering on leptons, this analysis is able to select events with high efficiency for a lower \MET threshold,  
$80$ GeV, than 
in the monojet search, which requires $\MET > 250$ GeV.  Two well-identified, isolated leptons of opposite sign and same flavor are required to have an invariant mass within 10 GeV of the Z mass.  
A shape-based fit is performed to the distribution of the transverse mass $m_T$.  The data are consistent with the background expectation, as shown for the dimuon channel in Figure~\ref{fig:VMETLep}.  

Limits are placed on the dark matter-nucleon scattering cross section for spin-dependent scenarios, as shown in Figure~\ref{fig:VMETLep}, and for spin-independent scenarios.  Limits are also set on the dark matter annihilation rate and on the effective coupling cut-off scale $\Lambda$. 
The most stringent limits on the scalar unparticle model parameters $\lambda$ and $\Lambda_U$ are obtained.

\section{X $\rightarrow$ H V}
Three recent searches look for a heavy resonance that decays to a Higgs boson and a vector boson V, either a W or Z.  Higgs decays to $\tau\tau$, $b\overline{b}$, and WW are covered by these searches.  
If the new resonance is massive, the H and V are boosted, and each may be reconstructed as a single merged jet.  The jets are identified based on quantities that characterize their substructure, such as the pruned jet mass.  The signal would appear as a peak in the distribution of the HV invariant mass. 

\subsection{X $\rightarrow$ H($\tau\tau$) + Z}
A search has been performed for a heavy resonance decaying to a Higgs boson in the $\tau\tau$ decay mode and a Z boson that decays hadronically~\cite{resonanceHTauTau}.  There are six channels analyzed to cover all $e$, $\mu$, and hadronic $\tau$ decay modes.  
In all channels, the signal region is defined by a pruned jet mass between 70--110 GeV and $\tau_{21} < 0.75$, where $\tau_{21}$ is the ratio of 2-subjettiness to 1-subjettiness.  The preselection cuts vary by channel, to cope with different backgrounds.  
In the $\tau_e\tau_e$, $\tau_\mu\tau_\mu$, and $\tau_e\tau_\mu$ channels, the dominant background is from Z($\tau\tau$) + jets events.  
In the $\tau_e\tau_h$ and $\tau_\mu\tau_h$ channels, the major backgrounds arise from 
Z + jets, W + jets, and $t\overline{t}$ events.  
The $\tau_h\tau_h$ channel faces backgrounds principally from QCD multijets events.  

\begin{figure}[!h]
\centering
\begin{minipage}{0.45\linewidth}
\centerline{\includegraphics[width=\linewidth]{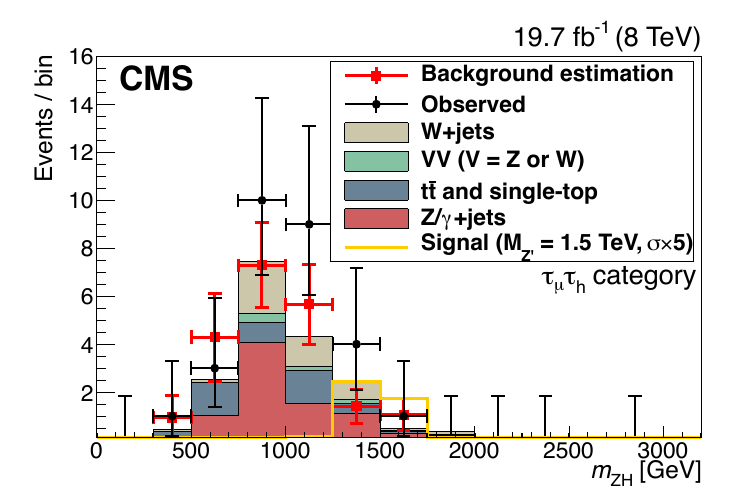}}
\end{minipage}
\hfill
\begin{minipage}{0.45\linewidth}
\centerline{\includegraphics[width=\linewidth]{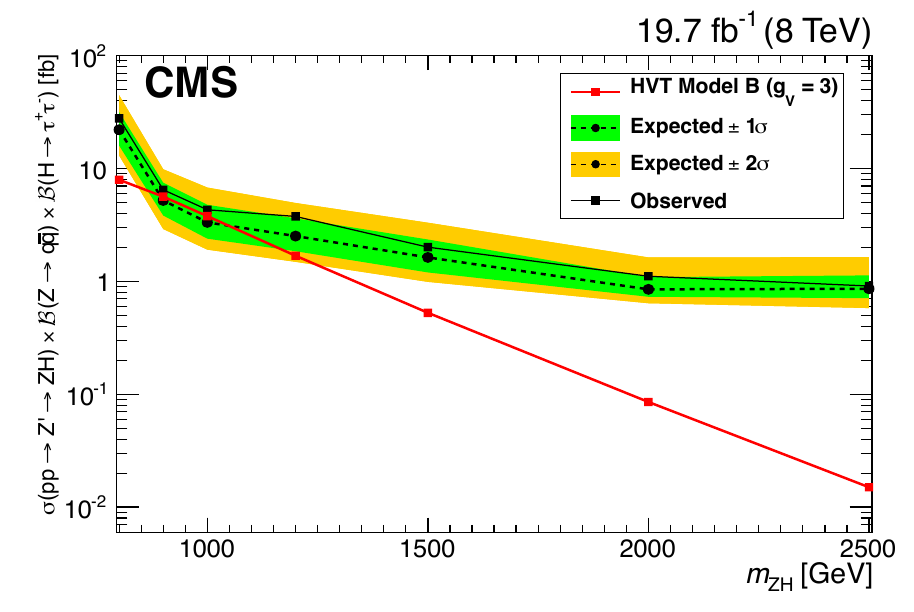}}  
\end{minipage}
\caption{ 
Left:  observed distribution of the ZH mass for the $\tau_\mu\tau_h$ channel, along with the data-driven estimate of the background, and the simulated background contributions.  
Right:  expected and observed limits on $\sigma(\rm{Z}') \times {\cal B}(\rm{Z}' \rightarrow \rm{ZH})$. 
}
\label{fig:resonanceHTauTau}
\end{figure}

In the six $\tau\tau$ decay channels and seven bins of the HZ invariant mass, the observation is compared with the expected background.  The HZ invariant mass distribution for the $\tau_\mu\tau_h$ channel is shown in Figure~\ref{fig:resonanceHTauTau}.  The backgrounds are estimated by extrapolating from sideband regions, using 
different techniques for the all-leptonic, semileptonic, and hadronic ττ decay channels.  
There are 0--13 observed events in each bin, and there are no significant deviations from the expected yield.
Limits at 95\% confidence of 0.9--27.8 fb are placed on the Z' cross section times branching fraction for $0.8 < m(\rm{HZ}) < 2.5$ TeV, as shown in Figure~\ref{fig:resonanceHTauTau}.

\subsection{X $\rightarrow$ H($b\overline{b}$) + W($\ell\nu$)}

A search has been performed for a heavy resonance decaying to a Higgs boson in the $b\overline{b}$ final state and a W boson that decays leptonically~\cite{resonanceHWLep}.  The electron and muon channels are considered.
The H($b\overline{b}$) is reconstructed from a single CA8 jet, with pruned jet mass of 110--135 GeV.  
A dedicated b-tagging technique is applied for the boosted H in which the jet is split into two subjets.  The dominant backgrounds in this search arise from W+jets and $t\overline{t}$ events.  

\begin{figure}[!h]
\centering
\begin{minipage}{0.45\linewidth}
\centerline{\includegraphics[width=\linewidth]{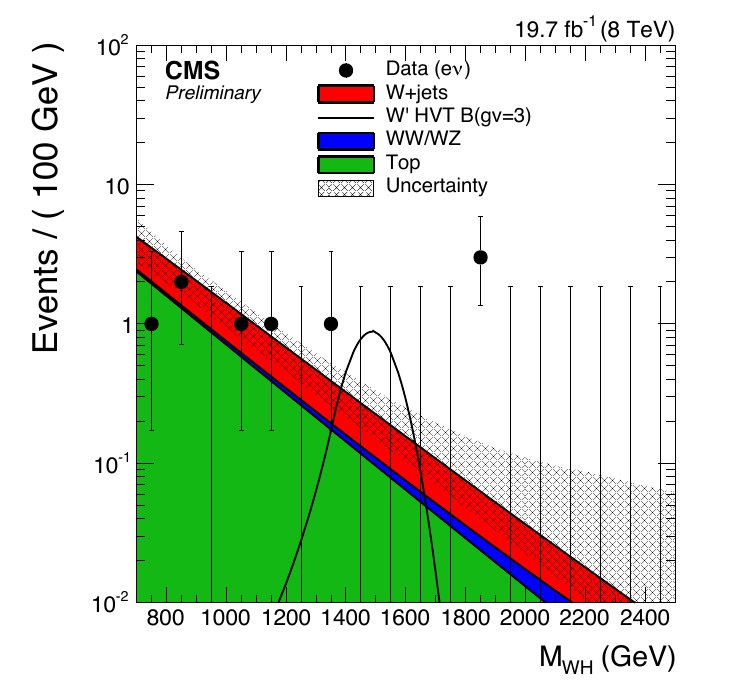}}
\end{minipage}
\hfill
\begin{minipage}{0.45\linewidth}
\centerline{\includegraphics[width=\linewidth]{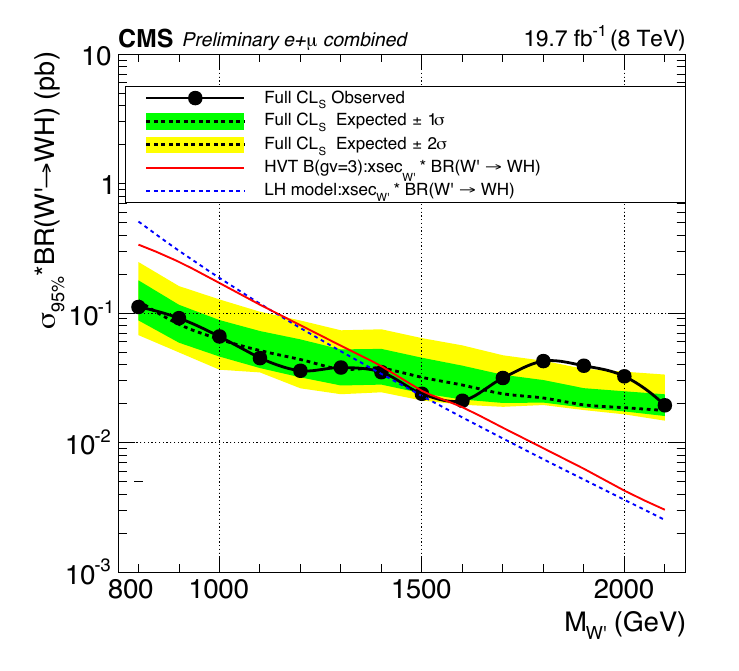}}  
\end{minipage}
\caption{ 
Left:  distribution of the HW invariant mass, for the electron channel. 
Right:  observed and expected upper limits on 
$\sigma(\rm{W}') \times {\cal B}(\rm{W}' \rightarrow \rm{WH})$. 
}
\label{fig:resonanceHWLep}
\end{figure}

The HW invariant mass distribution in data is compared with the prediction from data control regions, as shown for the electron channel in Figure~\ref{fig:resonanceHWLep}.  
Three events are observed in the electron channel with HW invariant mass between 1.8 and 1.9 TeV, corresponding to a 2.9 $\sigma$ local significance.  
Limits are placed on the W' cross section times branching ratio, as shown in Figure~\ref{fig:resonanceHWLep}.  
These limits exclude a W' with mass less than 1.4 TeV in the Little Higgs model and with mass less than 1.5 TeV in the Heavy Vector Triplet model.  

%
%

\subsection{X $\rightarrow$ H(WW,$b\overline{b}$) + W/Z}
\begin{figure}[!h]
\centering
\begin{minipage}{0.45\linewidth}
\centerline{\includegraphics[width=\linewidth]{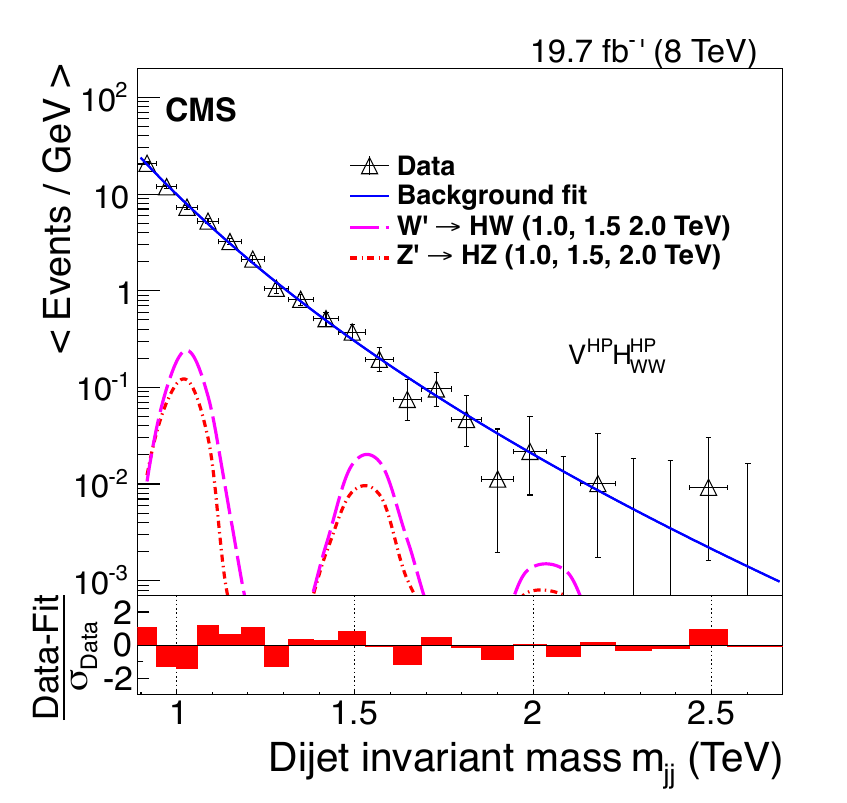}}  
\end{minipage}
\hfill
\begin{minipage}{0.45\linewidth}
\centerline{\includegraphics[width=\linewidth]{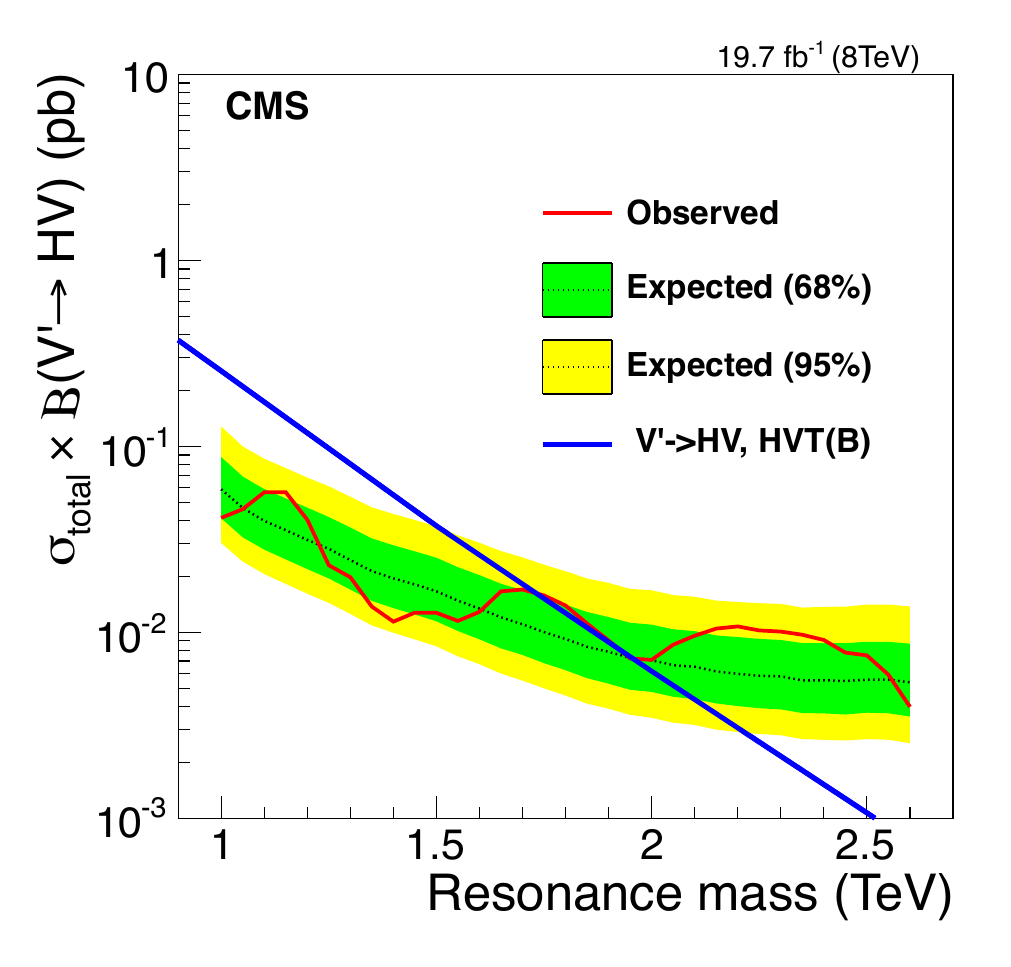}}  
\end{minipage}
\caption{ 
Left:  distribution of the dijet invariant mass for the category of high-purity H$\rightarrow$WW and high-purity V.  
Right:  observed and expected upper limits on $\sigma$(V'$\rightarrow$VH), obtained by combining the Z' and W' channels together.  
}
\label{fig:resonanceHWWbb}
\end{figure}

A search has been performed for a resonance decaying to a Higgs in the WW and \bbbar channels and a W or Z that decays hadronically~\cite{resonanceHWWbb}.  Boson jets are identified based on the pruned jet mass, dedicated subjet b-tagging, and N-subjettiness.  
There are five event categories, for the two Higgs decay modes, and for low-purity versus high-purity V or H tagging.  The background shape of the distribution of the HV invariant mass is modeled with a smoothly falling parametric function.  
The distribution for the category of high-purity H(WW) and V is shown in Figure~\ref{fig:resonanceHWWbb}.  
A common likelihood fit is performed to the HV invariant mass distributions in the five categories.  
In a Heavy Vector Triplet model, a Z' is excluded in the mass intervals [1.0, 1.1] and [1.3, 1.5] TeV, and a W' is excluded in the interval [1.0, 1.6] TeV.  
A mass degenerate W' plus Z' particle is excluded in the mass interval [1.0, 1.7] TeV, as shown in Figure~\ref{fig:resonanceHWWbb}.  

\section{Dijet resonances}
A search for exotic resonances that decay to pairs of jets has been conducted with 42 pb$^{-1}$ of proton-proton collisions collected at $\sqrt{s} = 13$ TeV~\cite{dijetResonances}.  
Such a resonance could be identified as a bump in the dijet invariant mass distribution.  
Events are collected with a trigger that requires the scalar sum of jet transverse momentum to be greater than 800 GeV.  
Jets are formed from Particle Flow candidates (charged/neutral hadrons, muons, electrons, and photons), and are clustered using an anti-kT algorithm with distance parameter 0.4.  
The jet momenta are corrected using calibration constants derived from simulations.  
The two leading jets must pass various identification criteria.  
"Wide jets" are reconstructed by adding jets within $\Delta R<1.1$, starting with the two leading jets as seeds.  
Various cross checks of basic jet and event quantities have been performed to ensure that good dijet events are selected.  

\begin{figure}[!h]
\centering
\begin{minipage}{0.75\linewidth}
\centerline{\includegraphics[width=\linewidth]{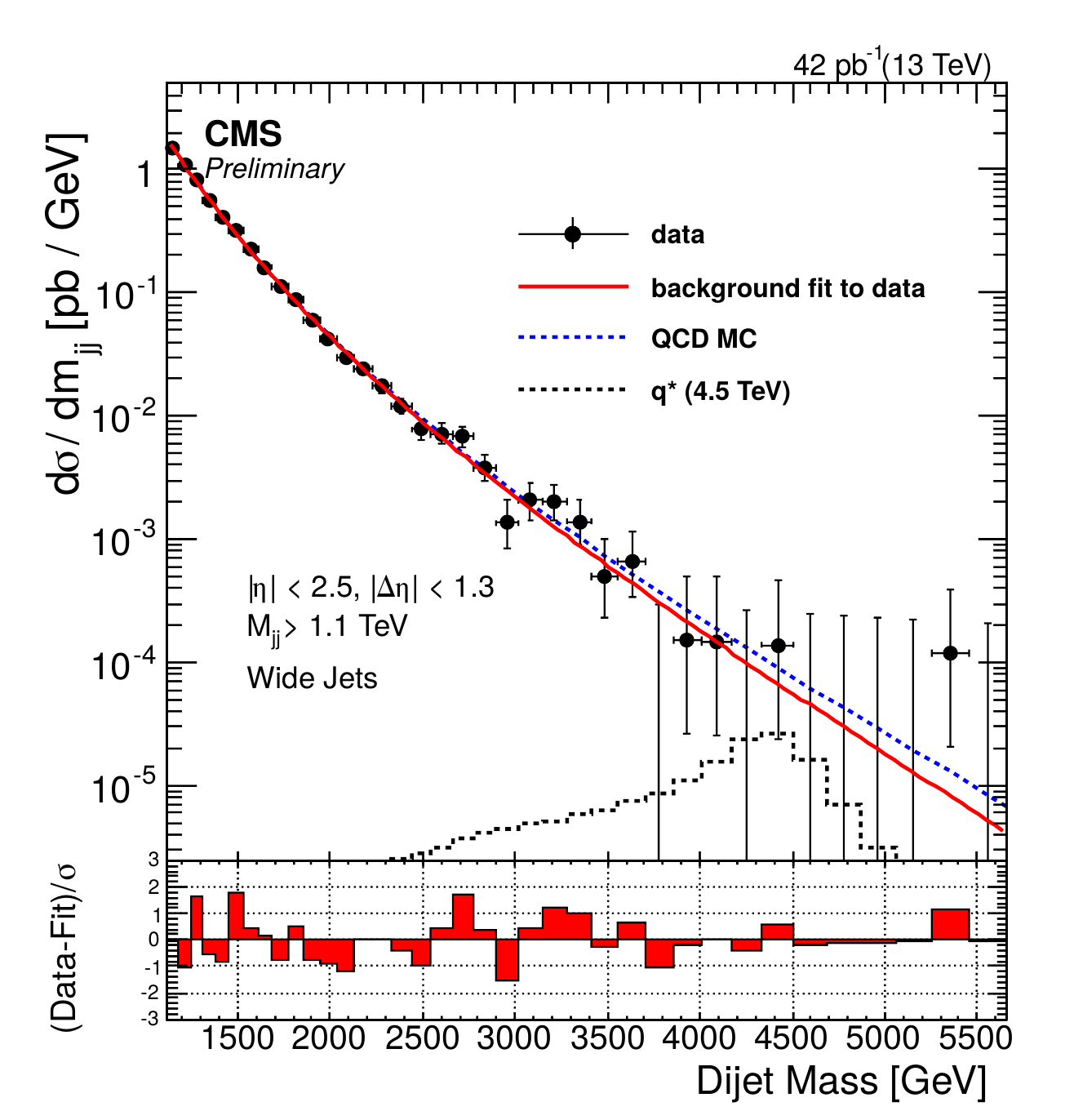}}  
\end{minipage}
\caption{ 
Distribution of the dijet invariant mass spectrum, calculated using wide jets, for observed data and the simulated contribution from QCD multijets events.  A fit using a smooth parametric function is also shown.
}
\label{fig:dijetRes}
\end{figure}
\begin{figure}[!h]
\centering
\begin{minipage}{0.45\linewidth}
\centerline{\includegraphics[width=\linewidth]{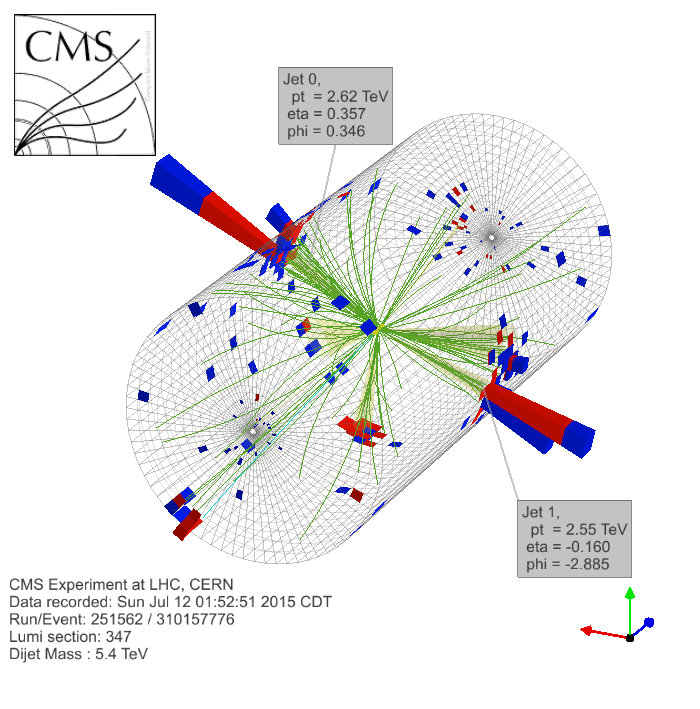}}  
\end{minipage}
\hfill
\begin{minipage}{0.45\linewidth}
\centerline{\includegraphics[width=\linewidth]{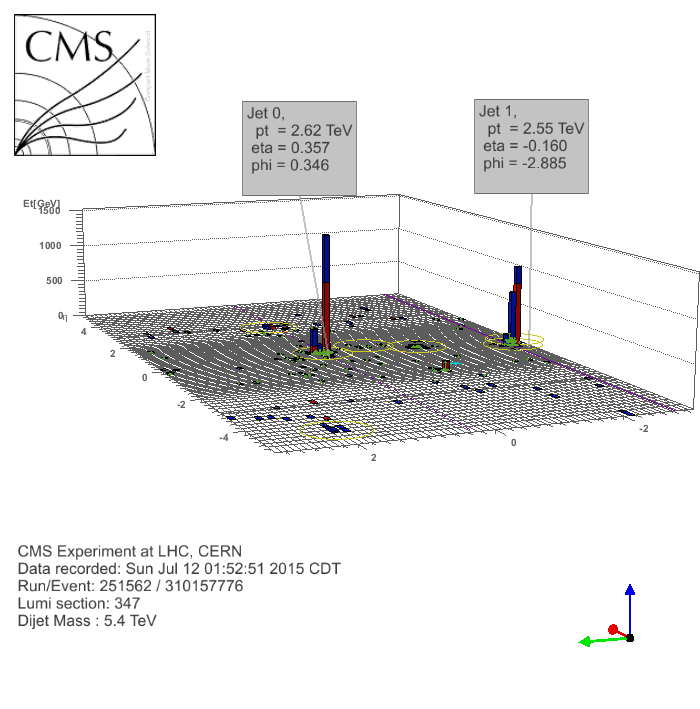}}  
\end{minipage}
\caption{ 
The three-dimensional view (left) and $\eta$-$\phi$ view (right) of the event with the largest dijet invariant mass, 5.4 TeV.  
}
\label{fig:dijetResEvtDisplay}
\end{figure}

A maximum likelihood fit is performed to the dijet mass distribution, using a four-parameter function, as shown in Figure~\ref{fig:dijetRes}.  
The fit $\chi^2$/ndof  is 25 / 34, consistent with the background-only hypothesis.  
The event with the largest dijet mass, 5.4 TeV, is shown in Figure~\ref{fig:dijetResEvtDisplay}.   

\section{Conclusion} 
Searches for new physics in the V+\MET final state have set new limits on dark matter-nucleon scattering cross sections.  
Searches for X$\rightarrow$HV have set lower limits on W', Z', and degenerate V' masses.
A search for dijet resonances has found the early Run 2 data to be well-described by a background-only parameterization function.  
CMS is poised to explore the new regime of $\sqrt{s} = 13$ TeV proton-proton collisions with a broad program of searches for exotic signatures of new physics.  

\Acknowledgments
I thank the many analysts who produced the results presented in this talk.  I also thank all of our CMS colleagues, the CERN accelerator division for the excellent operation of the LHC, and the many funding agencies that support our work.

\end{document}